# Full angular dependence of the spin Hall and ordinary magnetoresistance in epitaxial antiferromagnetic NiO(001)/Pt thin films

L. Baldrati[1], A. Ross[1,2], T. Niizeki[3], C. Schneider[1], R. Ramos[3], J. Cramer[1,2], O. Gomonay[1], M. Filianina[1,2], T. Savchenko[4], D. Heinze[1], A. Kleibert[4], E. Saitoh[3,5,6,7,8], J. Sinova[1], M. Kläui[1,2*]

[1]*Institute of Physics, Johannes Gutenberg-University Mainz, 55128 Mainz, Germany*

[2]*Graduate School of Excellence Materials Science in Mainz, 55128 Mainz, Germany*

[3]*Advanced Institute for Materials Research, Tohoku University, Sendai 980-8577, Japan*

[4]*Swiss Light Source, Paul Scherrer Institute, 5232 Villigen PSI, Switzerland*

[5]*Institute for Materials Research, Tohoku University, Sendai 980-8577, Japan*

[6]*Advanced Science Research Center, Japan Atomic Energy Agency, Tokai 319-1195, Japan*

[7]*Center for Spintronics Research Network, Tohoku University, Sendai 980-8577, Japan*

[8]*Department of Applied Physics, The University of Tokyo, Tokyo 113-8656, Japan*

**Electronic Mail: klaeui@uni-mainz.de*

## ABSTRACT

We report the observation of the three-dimensional angular dependence of the spin Hall magnetoresistance (SMR) in a bilayer of the epitaxial antiferromagnetic insulator NiO(001) and the heavy metal Pt, without any ferromagnetic element. The detected angular-dependent longitudinal and transverse magnetoresistances are measured by rotating the sample in magnetic fields up to 11 T, along three orthogonal planes (xy-, yz- and xz-rotation planes, where the z-axis is orthogonal to the sample plane). The total magnetoresistance has contributions arising from both the SMR and ordinary magnetoresistance. The onset of the SMR signal occurs between 1 and 3 T and no saturation is visible up to 11 T. The three-dimensional angular dependence of the SMR can be explained by a model considering the reversible field-induced redistribution of magnetostrictive antiferromagnetic S- and T-domains in the NiO(001), stemming from the competition between the Zeeman energy and the elastic clamping effect of the non-magnetic MgO substrate. From the observed SMR ratio, we estimate the spin mixing conductance at the NiO/Pt interface to be greater than $2x10^{14}$ $\Omega^{-1}$ $m^{-2}$. Our results demonstrate the possibility to electrically detect the Néel vector direction in stable NiO(001) thin films, for rotations in the xy- and xz- planes. Moreover, we show that a careful subtraction of the ordinary magnetoresistance contribution is crucial to correctly estimate the amplitude of the SMR.

**MANUSCRIPT**

## I. INTRODUCTION

Antiferromagnets (AFMs) are materials with compensated magnetic sublattices and a vanishingly small magnetization, presenting many technologically relevant properties: they are free from stray fields, robust against external magnetic perturbations and potentially operating at THz frequencies. [1,2] However, a reliable electrical reading of antiferromagnetically stored information is challenging, especially in antiferromagnetic insulators, [3] where a charge current cannot flow. A possible approach providing electrical information on the magnetic state of insulators is a new type of magnetoresistance, the spin Hall magnetoresistance (SMR), [4] which occurs in bilayers of a magnetic material and a heavy metal with high spin orbit coupling. The SMR is a magnetoresistance effect caused by the simultaneous interconversion between a charge current and a spin current, via the spin Hall effect (SHE) and inverse spin Hall effect (ISHE). [4–6] The SMR was first observed in ferrimagnets, but it occurs in antiferromagnets as well, [7,8] providing information on the orientation of the Néel vector, which otherwise can only be determined with X-ray magnetic linear dichroism (XMLD) at synchrotrons. [9–11] One can inject a spin current into a ferro(i)magnetic material (FM), exhibiting a net magnetization **M**, by driving a charge current $J_c$ into an adjacent HM layer along the *x*-direction. A spin accumulation $\boldsymbol{\mu}_s$, polarized along the *y*-direction (cf. Fig. 1b-d), is generated in the HM directly by the SHE and a spin current $J_s$ flows across the interface along the *z*-direction with an efficiency described by the spin mixing conductance $G_{\uparrow\downarrow}$. [6] This is a complex quantity, but in the case of yttrium iron garnet (YIG)/Pt it was shown that the real part of the spin-mixing conductance $G_r$ is one order of magnitude bigger than the imaginary part $G_i$. [6] We thus consider also for the NiO/Pt system the effect of the real part $G_r$ only, that is sufficient to explain our data. The spin polarization of the main diffusive component of $J_s$, arising from the real part of the spin mixing conductance $G_r$ only, is proportional to $\mathbf{m} \times (\mathbf{m} \times \boldsymbol{\mu}_S)$, [6] where $\boldsymbol{\mu}_s$ is the polarization of the spin accumulation and **m** is the normalized magnetization ($|\mathbf{m}| = 1$). The *y*-component of the spin polarization of $J_s$, proportional to ($1 - m_y^2$), generates a charge current along *x* (longitudinal resistance variation) via the ISHE, while the *x*-component, proportional to $m_x m_y$, generates a charge current along *y* (transverse resistance variation). Therefore, when $\mathbf{m} \parallel \boldsymbol{\mu}_s$, the spin current is reflected from the magnetic material [4,12] and the additional charge current induced by the ISHE leads to a lowered measured resistance. Then again, the spin current is absorbed by spin transfer torque if $\boldsymbol{\mu}_s \perp \mathbf{m}$, yielding a state of higher measured resistance in the HM Pt. The orientation of **m** can be controlled by an external magnetic field. In the simplest case, when the magnetization is aligned along **H**, as the angle α between the current and the magnetic field is changed, a $\cos^2(\alpha)$ modulation of the resistance, following the definition of α in Fig. 1b, is obtained (positive SMR by convention). [4] By contrast, in antiferromagnets the magnetic moments follow the normalized Néel vector **n** ($|\mathbf{n}| = 1$), therefore the main component of the polarization of the spin current flowing along z is proportional to $\mathbf{n} \times (\mathbf{n} \times \boldsymbol{\mu}_S)$, [4,7] i.e. the longitudinal SMR depends on $1 - n_y^2$, while the transverse SMR depends on $n_x n_y$. The orientation of the Néel vector can be also controlled by a sufficiently large magnetic field, however the configuration of lowest Zeeman energy is the one with $\mathbf{n} \perp \mathbf{H}$, [13] therefore the dependence of the SMR versus field is shifted by 90° as compared to ferromagnets, yielding an expected $\sin^2(\alpha)$ dependence (negative SMR).

Both positive and negative SMR have been observed in HM/antiferromagnetic insulator(AFMI)/FM trilayers, [14–17] where the spin backflow from the antiferromagnet and the ferromagnetic insulator (FMI) is detected. A non-monotonic behavior of the SMR as a function of the AFM thickness [14,16] as well as a change of sign as a function of temperature have been reported in AFMs [14–16,18] and in compensated ferrimagnets, [12] suggesting a competition between positive and negative contributions arising from the FMI and the AFMI, respectively. The SMR in bilayers of antiferromagnets/HM has been investigated only very recently. A positive SMR was claimed for bilayers of $SrMnO_3$/Pt [19] and $Cr_2O_3$/W, [20] possibly due to a small canted/uncompensated magnetization **M**. [21] During the preparation of this work, reports on a positive SMR in paramagnetic $Cr_2O_3$/Pt, negative SMR in $Cr_2O_3$/Ta [22,23] and a negative SMR in NiO(111)/Pt bilayers, [24,25] where the magnetic field rotates in the NiO(111) easy plane only, in agreement with theoretical predictions [7], were published. ~~during the preparation of this work.~~ However, at present, the three-dimensional angular dependence of the spin Hall magnetoresistance in an antiferromagnet has not yet been reported, nor has the SMR been shown in epitaxial NiO(001) thin films, the most stable NiO orientation, [26] relevant for applications.

In this paper, we analyze the occurrence of the SMR in a bilayer of Pt and epitaxial NiO(001) thin films, as revealed by angular-dependent magnetoresistance (ADMR) measurements along different rotation planes. We provide a theoretical model to describe the multidomain antiferromagnetic state of the NiO, based on the competition between the Zeeman energy and the destressing energy, stemming from the clamping effect of the substrate. [27] By using this model, in addition to the SMR theory described above, we can explain the full set of ADMR data acquired experimentally, after subtracting magnetoresistive contributions of the Pt not related to the NiO antiferromagnetic order. Finally, from the measured data, we gauge the efficiency of the spin transport in antiferromagnetic thin films, by ascertaining the spin mixing conductance of the NiO(001)/Pt interface. [6]

## II. THEORY

NiO is an easy-plane, collinear antiferromagnetic insulator (band gap of 4 eV) [28] with a bulk Néel temperature of 523 K, [29] representing an ideal model system to study the SMR in AFMs. Above the Néel temperature, NiO exhibits a cubic rock salt structure, while the unit cell contracts along one of the four <111> directions in the bulk antiferromagnetic state. [29] Within each of these four magnetostrictive domains (T-domains) the spins are confined to ferromagnetic $\{111\}$ planes, which in turn are antiferromagnetically ordered. [29–31] The spins are relatively free to rotate to point in one of the three $<11\bar{2}>$ directions, leading to three possible independent spin domains (S-domains) within each of the T-domains. Like T-domains, S-domains also possess different strains of magnetoelastic origin. The magnetic structure of NiO is shown schematically in Fig. 1a for a single S-domain, while all possible domains are listed in Table I and II. With the application of an external field, the main mechanism of spin realignment is the motion of the AFM domain walls, [13,29,31,32] which was reported to occur between <1 T and 7 T for bulk crystals. [13,32–34]

As mentioned above, the SMR depends on the relative alignment of the Néel vector orientation **n** and the current direction. [7,25] In a multidomain sample, the resistance variation depends on the average Néel vector of the domain structure. Therefore, the longitudinal (transverse) SMR along (perpendicular to) the current direction can be modeled as:

$$\frac{\Delta R_{xx}}{\bar{R}} = -\frac{\Delta\rho}{\rho}\langle n_y^2\rangle, \quad \frac{\Delta R_{xy}}{\bar{R}}\frac{l}{w} = 2\frac{\Delta\rho}{\rho}\langle n_x n_y\rangle, \quad (1)$$

where the $x$ axis is parallel to the current (see Fig. 1a-d), $\Delta\rho/\rho$ represents the SMR coefficient, and $\langle\ldots\rangle$ means averaging over the domain structure with the additional limitations discussed below. Note that these expressions are obtained under the hypothesis of negligible imaginary part of the spin mixing conductance. As pointed out above, the application of an external magnetic field $\mathbf{H}$ to the NiO sample induces a reversible redistribution of the AFM domains, which affects the SMR. Our model assumes that the equilibrium domain structure of the NiO is defined by the competition between magnetoelastic effects and the external magnetic field. [3,21,25] The reversible redistribution of the magnetic domains by external fields can be explained by the destressing fields, similar to the demagnetizing fields in ferromagnets, which originate from long-range elastic forces. [3] In particular, in NiO, which shows pronounced magnetoelastic coupling, [35] the domains with different orientations of the Néel vector $\mathbf{n}^{(0)}$ are strained in different directions, $u_{jk}^{(0)} \propto n_j^{(0)} n_k^{(0)}$ (see Table II). The strain of each homogeneously deformed region $\hat{u}^{(0)}$ alone is incompatible with the non-deformed nonmagnetic substrate (clamping effect). Thus, a monodomain state produces macroscopic mechanical stresses and is not energetically favorable. By contrast, the formation of a multidomain state allows the system to reduce the average strain to zero, since shear stresses, having opposite signs in different domains, can be relaxed. We model this effect by introducing the destressing energy [3,27] (per unit volume), whose structure corresponds to the tetragonal symmetry of the sample (NiO and the substrate). If the $z$ axis is perpendicular to the plane of the sample, we define:

$$
\begin{aligned}
&E_{\text{destr}} = \frac{1}{2} H_{\text{destr}}^{(\alpha)} M_s \left[ \langle n_x^2 - n_y^2 \rangle^2 + 4 \langle n_x n_y \rangle^2 \right] \\
&+ H_{\text{destr}}^{(\beta)} M_s \left[ \langle n_x^2 - n_z^2 \rangle^2 + \langle n_y^2 - n_z^2 \rangle^2 \right] \\
&+ 2 H_{\text{destr}}^{\perp} M_s \left[ \langle n_x n_z \rangle^2 + \langle n_y n_z \rangle^2 \right]
\end{aligned}
\tag{2}
$$

where $M_s$ is the saturation magnetization, $H_{\text{destr}}^{(\alpha)}$, $H_{\text{destr}}^{(\beta)}$, and $H_{\text{destr}}^{\perp}$ are the phenomenological constants depending on the NiO surface and the interface with the MgO substrate. Within this model and assuming perfectly movable domain walls, the domain structure is described by the relative fractions $\xi_j$, $j = 1, \ldots, N$ of the different domains, which can be considered as variables. The number of variables depends on the number of possible domains *N*. In our experiments we assume NiO(001) samples with all possible types of T- and S- domains (see Table I), which means that *N* = 12 and the domain structure provides 11 additional degrees of freedom ( $\sum_{j=1}^{N} \xi_j = 1$). This is in contrast to Ref. [25], where the authors considered a NiO(111) film, whose structure favours the formation of only one T-domain (T1 in our notation). In that case, the domain structure consists of S-domains only, i.e. *N* = 3, which gives two additional degrees of freedom. As a result of the more complicated domain structure of NiO(001), we use the full set of parameters for modelling the destressing fields, while in Ref. [25] the two constants $H^{(\beta)}_{\rm destr}$ and $H^{\perp}_{\rm destr}$ could be set to zero. We derive the model below for a Pt Hall bar aligned along the [100] direction, for a NiO(001) sample. Note that, thanks to symmetry, all the results hold for a Hall bar aligned along the [010] direction.

In the absence of external fields, the destressing energy (2) is minimum when $\langle n_j^2 - n_k^2\rangle = \langle n_j n_k\rangle = 0$, which can be achieved in case of uniform distribution of the domains ( $\xi_j = 1/12$ ). However, the application of an external magnetic field removes the degeneracy of the domains by introducing the additional energy contribution (Zeeman energy)

$$E_{\rm Zee} = \frac{M_s}{2H_{\rm ex}}\left\langle (\mathbf{H}\cdot\mathbf{n})^2 \right\rangle, \qquad (3)$$

where the field $H_{ex}$ parametrizes the exchange coupling between the magnetic sublattices. The Zeeman energy can be minimized by the coherent rotation of the Néel vectors within the domains and/or by the redistribution of the domain structure (variation of $\xi_j$). If the pinning strength of the domain walls is small as compared to the spin-flop field, the mechanism based on the domain wall motion is preferable. In this case, the effect of the external magnetic field can be fully or partially screened by the domain redistribution (for the details on the screening mechanism see Ref. [25]). In NiO(001) samples a variety of T- and S- domains is sufficient to provide full screening of the effect of the magnetic field. In this case, the orientation of the Néel vectors inside the domains is defined by the magnetocrystalline anisotropy only (see Table II) and the domain fractions can be calculated by the minimization of the sample energy $E_{\rm destr} + E_{\rm Zee}$ with respect to $\xi_j$, or any set of independent linear combinations. The stress/strain tensor has five independent components corresponding to shear strains, which can be relaxed by the redistribution of the 12 types of domains in different ways. Therefore, we here use five independent combinations generated by $\langle n_x^2 - n_y^2\rangle$, $\langle 2n_z^2 - n_x^2 - n_y^2\rangle$, $\langle n_x n_y\rangle$, $\langle n_y n_z\rangle$, and $\langle n_x n_z\rangle$ ($\langle n_x^2+n_y^2+n_z^2\rangle=1$), which vary from zero (if $\mathbf{H}=0$) to the limiting values in a single-domain state:

$$|\langle n_x^2-n_y^2\rangle|\le 1/2,\quad -1/2\le\langle 2n_z^2-n_x^2-n_y^2\rangle\le 1,\quad |\langle n_j n_k\rangle|\le 1/3. \qquad (4)$$

By minimizing $E_{\mathrm{destr}}+E_{\mathrm{Zee}}$ with respect to all independent variables one can obtain the domain distribution as a function of the external magnetic field within the bulk sample. However, the redistribution of the domains under the action of the magnetic field proceeds through the motion of the domain walls. By looking at the x-ray magnetic linear dichroism (XMLD) image shown in Fig. 1e, discussed below in the Experimental section, one notices that the antiferromagnetic domain walls are oriented at 45° with respect to the crystallographic directions $(100)$ and $(010$ $)$, implying that the domain wall orientation in our NiO films coincides with the crystallographic planes $\{1\bar{1}0\}$. We thus exclude the contribution from the $<01\bar{1}>$ domain walls not seen in our XMLD analysis, by minimizing the energy with the additional restriction $\langle n_y^2-n_z^2\rangle=0$. By calculating the values of $\langle n_y^2\rangle$ and $\langle n_x n_y\rangle$, and by substituting them into Eq. (1), we obtain the angular dependence of the SMR for all the three orthogonal plane scans, depicted in Figs. 1a-d, as follows:

$\alpha$-scan (*xy*-plane):

$$\frac{\Delta R_{xx}^{(\alpha)}}{\bar{R}}=\frac{\Delta\rho}{\rho}\frac{H^2}{4H_{\mathrm{ex}}(H_{\mathrm{destr}}^{(\alpha)}+H_{\mathrm{destr}}^{(\beta)})}\sin^2\alpha,\quad \frac{\Delta R_{xy}^{(\alpha)}}{\bar{R}}\frac{l}{w}=-\frac{\Delta\rho}{\rho}\frac{H^2}{8H_{\mathrm{ex}}H_{\mathrm{destr}}^{(\alpha)}}\sin 2\alpha\,, \qquad (5)$$

$\beta$-scan (*yz*-plane):

$$\Delta R_{xx}^{(\beta)}/\bar{R}=0,\quad \Delta R_{xy}^{(\beta)}/\bar{R}=0, (6)$$

$\gamma$-scan (*xz*-plane):

$$\frac{\Delta R_{xx}^{(\gamma)}}{\bar{R}}=-\frac{\Delta\rho}{\rho}\frac{H^2}{4H_{\mathrm{ex}}(H_{\mathrm{destr}}^{(\alpha)}+H_{\mathrm{destr}}^{(\beta)})}\sin^2\gamma,\quad \frac{\Delta R_{xy}^{(\gamma)}}{\bar{R}}=0\,. \qquad (7)$$

Where *l/w* is the geometrical ratio length/width of the Hall bar, necessary to compare transverse and longitudinal measurements. Although relations (5)-(7) contain several phenomenological parameters that cannot be easily determined a priori, they predict the angular dependencies and relations between the amplitude of the SMR for different plane scans. In particular, Eq. (5) predicts a negative sign of the SMR within the *xy*-plane scan, i.e. the highest (lowest) Pt resistance is found for $\alpha = 90^\circ$ ($\alpha = 0^\circ,\ 180^\circ$). This is in agreement with previous theoretical predictions on antiferromagnets and previous experimental results on NiO. [7,14–16,24,25] In a similar way, Eq. (7) predicts the lowest Pt resistance for $\gamma = 90^\circ$, which, as in the case of the $\alpha$-scan, corresponds to the parallel alignment of $\mathbf{H}$ and the current. For the $\beta$-scan, in which $\mathbf{H}$ is always perpendicular to the current, the Pt resistance does not depend on the angle because of the restriction discussed above. The transverse resistance variation $\Delta R_{xy}$ vanishes in both out-of-plane scans $\beta$ and $\gamma$. Equations (5)-(7) also show that the amplitudes of the SMR signal,

$$\Delta_{xx}^{(\alpha)} \equiv \Delta_{xx}^{(\gamma)} \equiv \frac{\Delta\rho}{\rho}\frac{H^2}{4H_{\text{ex}}(H_{\text{destr}}^{(\alpha)}+H_{\text{destr}}^{(\beta)})}, \quad \Delta_{xy}^{(\alpha)} \equiv \frac{\Delta\rho}{\rho}\frac{H^2}{4H_{\text{ex}}H_{\text{destr}}^{(\alpha)}}, \qquad (8)$$

grow with the magnetic field as $\propto H^2$. Equations (5)-(7) are valid whilst the sample is in a multidomain state and the domain fractions $\xi_j$ can be considered as variables. The field at which a monodomain state is formed, $H_{\text{MD}}$, can be obtained from the conditions that the variables take their limiting values corresponding to a single domain, say T1S1 (see e.g. Eq. (4) and Table II). Presupposing consistency of these conditions obtained for different variables, we obtain additional relations between the destressing fields: $H_{\text{destr}}^{(\alpha)} = H_{\text{destr}}^{(\perp)} = 3H_{\text{destr}}^{(\beta)}$. Under these restrictions, the value of $H_{\text{MD}}^{(\alpha)} = \sqrt{8H_{\text{ex}}H_{\text{destr}}^{(\beta)}}$ is the same for all types of plane scans. The obtained relation between the destressing fields $H_{\text{destr}}^{(\alpha)} = 3H_{\text{destr}}^{(\beta)}$ imposes an additional relation between the SMR amplitudes for different plane scans, which can be directly compared with the experiments:

$$\Delta_{xy}^{(\alpha)} = \frac{4}{3}\Delta_{xx}^{(\alpha)},\ \Delta_{xx}^{(\alpha)} = \Delta_{xx}^{(\gamma)},\ \Delta_{xx}^{(\beta)} = 0 \qquad (9)$$

## III. EXPERIMENTAL

The investigated NiO thin films of 90 nm thickness were grown on a (001)-oriented MgO substrate by means of radio frequency (RF) reactive magnetron sputtering. Epitaxial growth of NiO stems from the small lattice mismatch of ~1% between NiO ($a_{NiO}$ = 4.176 Å) and MgO ($a_{MgO}$ = 4.212 Å), both exhibiting a rock salt structure. The samples were grown in a sputtering system QAM4 from ULVAC, at a base pressure of $10^{-5}$ Pa. The MgO substrates were pre-annealed in vacuum at 800 °C for 2h. Subsequently, the NiO films were deposited from a Ni target in a steady Argon (15 sccm) and Oxygen (4.2 sccm) flow at 450 °C. Finally, a Pt top layer ($d_{Pt}$ = 7.5 nm) for the SMR experiment was deposited in-situ, after cooling down the samples to room temperature in vacuum. Epitaxial growth of the NiO(001) films was verified by means of X-ray diffraction and reciprocal space mapping, as discussed in detail elsewhere. [36]

The antiferromagnetic structure of a MgO//NiO(90 nm)/Pt(7.5 nm) sample was imaged by combing x-ray photoemission electron microscopy with the x-ray magnetic linear dichroism (XMLD) effect, after thinning the Pt layer by Ar ion etching. The images were acquired at the Ni $L_2$ edge using x-rays with linear vertical polarization and an energy of 871.0 eV and 869.6 eV [10] and calculating the XMLD as the difference of the images divided by the sum, while the x-ray beam was directed approximately along the (110) direction and at a grazing angle of 16° with respect to the surface of the sample. At $\mu_0 H$ = 0 T, our samples exhibit antiferromagnetic domains as shown in in Fig. 1e, demonstrating the high quality of our NiO films and the presence of displaceable domain walls, oriented at 45° with respect to the (100) or (010) sample directions. Note that the XMLD contrast vanishes at 533 K, as expected with respect to a bulk Néel temperature of T = 523 K, and no ferromagnetic contrast was seen in XMCD-PEEM measurements (see Supplementary information). Image distortions do not make

XMLD-PEEM imaging possible at high applied magnetic fields so that one cannot probe the SMR by direct imaging.

To check the validity of the model, the NiO/Pt bilayers were patterned into Hall bars using electron beam lithography and ion beam etching. The Hall bar geometry is shown in Fig. 1f. The samples were installed in a variable temperature insert cryostat with a superconducting magnet and static magnetic fields of up to 11 T were applied. The sample was rotated through the *xy*-, *yz*- and *xz*-planes (Figs. 1b-d), by means of a piezoelectric rotator ANRv51RES from Attocube, operated in an open loop. While rotating the sample, a charge current of density $9 \times 10^8$ A $m^{-2}$ from a Keithley 2400 was passed through the Hall bar and the transverse and longitudinal resistance were simultaneously captured with two Keithley 2182A nanovoltmeters. The sample was rotated clockwise by 180° degrees, then the field was reversed, and the sample was rotated counter-clockwise to the initial position. The longitudinal and transverse resistance were calculated as the average sum and difference, respectively, of positive and negative currents. Drifts in the resistance occurring during the reversal of the field were corrected for by summing a constant offset to the backward part, to ensure the continuity of the resistance. Linear drifts due to a non-perfect temperature stabilization were corrected when detected. The sample temperature was maintained at 199.46 ± 0.02 K for all measurements, as probed by a Cernox-1050 sensor.

## IV. RESULTS AND DISCUSSION

The full angular dependence of the MR amplitude for external fields $\mu_0\mathbf{H}$ up to 11 T and rotated in the *xy*-, *yz*- and *xz*-planes are shown in Figs. 2a-f. Note that a component of magnetoresistance appears in all the three planes. The SMR is expected in our model to be observable in the *xy*- and *xz*-planes scans only, based on the type of domain walls that were assumed, as discussed in the theory section. In YIG/NiO/Pt trilayers, the NiO has been shown to suppress the magnetic proximity effect in the Pt layer, [14] therefore we can exclude the presence of standard anisotropic magnetoresistance (AMR). Our data cannot be explained by the Hanle magnetoresistance since a maximum of the Hanle resistance was measured in the *xy*-plane scan for α = 0° in the YIG/Pt system, [37] while we observe a minimum for the same angle and field of 11 T. At this high field and Pt thickness, however, a contribution of ordinary magnetoresistance (OMR) can arise in Pt. [38] In order to test the presence of other types of magnetoresistance not related to the antiferromagnetic order, we grew and patterned a MgO//Pt(7.5 nm) control sample. The deposition condition and the device geometry were the same as for the MgO//NiO/Pt sample. The resistances are identical within the error set by the thickness calibration for the same nominal Pt thickness. The comparison between the MgO//Pt and the MgO//NiO/Pt, in the same experimental conditions at 11 T, is shown along all the three possible plane scans in Fig. 3a-f. As can be seen, the longitudinal magnetoresistance of the MgO/Pt is significantly different than the one of the MgO//NiO/Pt sample used for this study. The relative resistance variation in the sample without NiO is 27% of the one detected in the sample including NiO in the *xy*-plane scan and 35% in the *xz*- plane scan. However, the *yz*-plane scan is similar in the two samples, by which we conclude that this plane scan is dominated by the ordinary magnetoresistance. By contrast, the Hall effect, shown in panels 3d,f has the same magnitude in both samples, as expected for identical Pt thickness and Hall bar geometry, highlighting the comparability of the samples. Therefore, we can identify the SMR contribution as the additional fraction of magnetoresistance in the NiO/Pt sample compared to the control MgO/Pt sample. To quantitatively understand why the OMR has different amplitudes in all three planes, one has to consider in detail the geometry of the sample and the band structure of the Pt. We will just give qualitative arguments here, since a more detailed description is beyond the scope of the present paper. Note that a report highlighting the importance of the OMR assumed no contributions in the *xy*-plane scan, [38] as expected for a 2-D system, while we show here that these are not negligible in our bilayers. In a 3-D rod of material, where the current flows along the rod axis, the direction of the magnetic field (transverse/longitudinal) with respect to the current defines the ordinary magnetoresistance. Therefore, since both transverse directions *y* and *z* are equivalent, one observes resistance variations in the *xz*- and *xy*-planes scan, and no resistance variation in the *yz*-plane scan. By contrast, in a 2-D system, where the motion of the electrons is confined in the *z* directions, only the magnetic field component along *z* is important for the OMR since the electrons moves only in the *xy*-plane scan and the magnetic field component parallel to *y* has no effect. [38] In the 2-D case, equal resistance variation is expected in the *yz*- and *xz*-planes scans, while no variation occurs in the *xy*-plane scan. A Pt layer of 7 nm thickness is not a true 2-D system since the motion of the electrons is not perfectly confined in the *z* direction. In this case, which is the most common case in experiments and unlike a true 2-D system, a resistance variation arising from OMR is detected in all the *xy*-, *yz*- and *xz*-scans.

The longitudinal ($\Delta R_{xx}$) and transverse ($\Delta R_{xy}$) resistance variations of the MgO//NiO/Pt sample, measured in the *xy*-plane scan, were respectively fitted with a $\Delta R*\sin^2(\alpha + \alpha_{0,xx})$ and $\Delta R*\sin^2(\alpha + \alpha_{0,xy})$ functions (the fit is shown as continuous lines in Fig. 2a-b), where $\alpha_{0,xx}$ and $\alpha_{0,xy}$ are constants. The small differences of the offsets $\alpha_{0,xx} = 6° \pm 2°$ and $\alpha_{0,xy} = -40° \pm 2°$ to the expected values of 0° and -45°, respectively, are likely given by a non-corrected misalignment of the direction of the current, while their difference $\alpha_{0,xx} - \alpha_{0,xy} = 46 \pm 3°$ is in agreement to what is expected for SMR. The data shows a negative sign of the SMR within the *xy*-plane, i.e. the longitudinal resistance of the Pt is highest when the external field $\mathbf{H} \parallel \boldsymbol{\mu}_s$, and minimized for $\mathbf{H} \perp \boldsymbol{\mu}_s$, as predicted by Eq. (5). This implies that the SMR in our system results from a component of the Néel vector perpendicular to the field and supports the validity of our model. For the *yz*- and *xz*-planes scans, the transverse signal is dominated by the ordinary Hall effect of the Pt top layer, which is linear with the field and follows a $-\cos(\theta + \theta_{0,Hall})$ functional angular dependence. By subtracting the Hall-effect related $\cos(\theta + \theta_{0,Hall})$ angular dependence, we obtain a vanishing residual transverse magnetoresistance, as shown in Figs. 2d,f, in agreement with Eqs. (6,7). By correcting the offset in order to set $\theta_{0,Hall} = 0°$ and by applying the same correction to the transverse and longitudinal data, we find that the longitudinal resistance modulation in the *xz*-plane scan is $\Delta R^{*}\sin^2(\gamma + \gamma_{0,xx})$ with $\gamma_{0,xx} = 91 \pm 2°$, in agreement with Eq. (7).

As pointed out above, a contribution from ordinary magnetoresistance is present in Pt at high magnetic fields. This contribution is less than one third of the total magnetoresistance measured in the NiO/Pt at 11 T in the *xy*- and *xz*-planes scans, as determined by the comparison to the MgO/Pt control sample, but in the *yz*-plane scan the total MR is mainly arising from the OMR, while the SMR contribution is negligible considering an interval of 1 standard deviations around the measured value. Based on the ratio between OMR and SMR at 11 T in the three planes, we subtracted the same relative contribution from the field dependence. We can therefore fit the magnetoresistance data with our model after subtracting the OMR contribution. The SMR amplitudes in the three planes are shown in Fig. 4, together with the fit from the model. The longitudinal and transverse SMR amplitudes increase quadratically with the field and show no saturation as expected from Eq. (8). The transverse resistance $\Delta R_{xy}$ value has been multiplied by the geometric scaling factor of 8.67, calculated as the ratio between the length and width of the Hall bar. Note that, according to Eq. (5), the longitudinal and transverse components are shifted by 45°, which agrees well with the measured data shown in Figs. 2a,b. The phases of all plane scans agree with the values predicted in Eqs. (5-7). At 11 T, the longitudinal SMR amplitudes are $\Delta R_{xx}^{(\alpha)} / \overline{R} = \left(5.2 \pm 0.4\right) x10^{-5}$, $\Delta R_{xx}^{(\beta)} / \overline{R} = \left(5 \pm 9\right) x10^{-6}$, $\Delta R_{xx}^{(\gamma)} / \overline{R} = \left(7.7 \pm 1.0\right) x10^{-5}$, in the *xy*-, *yz*- and *xz*-plane scans, respectively, with the error being estimated by performing several times the same measurement. The transverse SMR amplitude, then again, is $\Delta R_{xy}^{(\alpha)} \times (l / w) / \overline{R} = \left(5.9 \pm 0.4\right) x10^{-5}$ in the *xy*-plane scan and zero (within the error) elsewhere, where $l / w = 8.67$ is the ratio length/width of the Hall bar. The absence of a detectable transverse SMR in the *xz*- and *yz*- plane scans, arising from the imaginary part of the spin-mixing conductance $G_i$, justifies the approximation of considering the real part $G_r$ only in the modeling.

The relative amplitudes of the longitudinal resistance variation in the different planes qualitatively agree with Eq. (9), which predicts no resistance variation in the *yz*-plane scan and a transverse resistance variation larger than the longitudinal resistance variation in the *xy*-plane scan. The different amplitudes between the *xy*- and *xz*-plane scans are not explained by our simplified model. This discrepancy might be due to defects affecting the domain wall motion, such as pinning defects, which have been neglected in the theory presented above. The study of these defects requires additional assumptions and experiments, which go beyond the aim of this paper.

In our experiments, no saturation of the signal up to 11 T is observed. The absence of an incipient saturation at 11 T is in agreement with the saturation field of 13 T reported in NiO(111) thin films, [25] considering that the NiO(001) orientation is even less favorable for the field-induced reorientation. From the measured SMR ratio $\Delta\rho/\rho$ at 11 T, calculated as the difference between the total MR ratio and the OMR in the control sample, we can set a lower limit for the spin mixing conductance by using the formula from the model based on spin diffusion theory by Chen et al.: [6]

$$\frac{\Delta\rho}{\rho} = \theta_{SH}^2 \frac{\lambda}{d_{Pt}} Re\left(\frac{2\lambda G_{\uparrow\downarrow}\tanh^2\left(\frac{d_{Pt}}{2\lambda}\right)}{\sigma + 2\lambda G_{\uparrow\downarrow}\coth\left(\frac{d_{Pt}}{\lambda}\right)}\right) \quad (10)$$

where $\sigma$ is the measured Pt conductivity ($5.5\times10^6\ \Omega^{-1}\ m^{-1}$), $\lambda$ is the spin relaxation length in the Pt (3.5 nm), [39] $\theta_{SH}$ is the spin Hall angle of the Pt (0.03), [39] and $d_{Pt}$ is the Pt thickness (7.5 nm). The spin relaxation length and the spin Hall angle were estimated by using the conductivity of the Pt and the values in Ref. [39]. By using these values, we obtain the lower limit for the real part of the spin mixing conductance at the NiO/Pt interface to be $G_{\uparrow\downarrow} \sim 2\times10^{14}\ \Omega^{-1}\ m^{-2}$. This value is comparable to what was obtained in YIG/Pt [4] and it fits well with the order of magnitude of recent theoretical predictions, [7] while it is lower than what was reported by Han et al. [19] in Pt/$SrMnO_3$. However, from the data by Hoogeboom et al. [24] in single crystals of NiO(111)/Pt, where a strong dependence on the interface treatment was found, and considering that 11 T are not sufficient to saturate the signal, we expect that the spin mixing conductance can be further optimized.[1]

## V. CONCLUSION

In conclusion, we report a full angular dependence study of the spin Hall magnetoresistance in NiO(001)/Pt epitaxial thin films grown on MgO(001) substrates. The measured symmetry of the spin Hall magnetoresistance is consistent with theoretical expectations [7] and agrees qualitatively with our model based on the redistribution of antiferromagnetic T- and S-domains. When the NiO(111) easy plane is not parallel to the surface of the NiO film, the effect of the magnetic field is non-trivial. We take into account the redistribution of different S- and T- domains, as resulting from the competition between the Zeeman and the destressing energy, for a thin film experiencing a clamping effect from the substrate. The model explains the quadratic field dependence of the SMR ratio, quantitatively predicts the phase of the SMR in all planes and qualitatively predicts relations between the amplitudes of the SMR in the different planes. A careful subtraction of the previously neglected ordinary magnetoresistance in the Pt layer is crucial to correctly estimate the SMR, which must be considered in view of applications and of using transport measurements to identify the Néel vector orientation of the antiferromagnet. From the observed SMR ratio, we estimate a spin mixing conductance greater than $2\mathrm{x}10^{14}\ \Omega^{-1}\ \mathrm{m}^{-2}$, which is close to the one observed in YIG/Pt. [4] Our results demonstrate that the SMR in epitaxial thin films is measurable even in the presence of a clamping effect from a substrate and for crystal orientations not favorable for the magnetic moment reorientation in an easy plane, making another step towards the all-electrical detection of the magnetic moments in this class of materials.

## ACKNOWLEDGEMENTS

[1] The spin mixing conductance of $10^{18}\ \Omega^{-1}\ \mathrm{m}^{-2}$, claimed in single crystalline NiO(111)/Pt, [24] is overestimated due to a miscalculation (G. R. Hoogeboom, private communication).

This work was supported by Deutsche Forschungsgemeinschaft (DFG) SPP 1538 “Spin Caloric Transport,” the Graduate School of Excellence Materials Science in Mainz (MAINZ), the EU project INSPIN (FP7-ICT-2013-X 612759) and ASPIN (EU FET Open RIA Grant No. 766566). The authors acknowledge the support of SpinNet (DAAD Spintronics network, project number 56268455), MaHoJeRo (DAAD Spintronics network, project number 57334897), the SNSF (project number 200021_160186), and the DFG (SFB TRR 173 SPIN+X). O.G. and J.S. acknowledge the Alexander von Humboldt Foundation and the ERC Synergy Grant SC2 (No. 610115). This work was supported by ERATO “Spin Quantum Rectification Project” (Grant No. JPMJER1402) from JST and Grant-in-Aid for Scientific Research on Innovative Area, “Nano Spin Conversion Science” (Grant No. JP26103005) from JSPS KAKENHI, Japan, and the NEC Corporation. Part of this work was performed at the Surface/Interface:Microscopy (SIM) beamline of the Swiss Light Source, Paul Scherrer Institut, Villigen, Switzerland. M.K. thanks ICC-IMR at Tohoku University for their hospitality during a visiting researcher stay at the Institute for Materials Research. The authors acknowledge useful scientific discussion with G. E. W. Bauer, R. Lebrun, D.-S. Han and K. Lee, as well as skillful technical support from J. Henrizi.

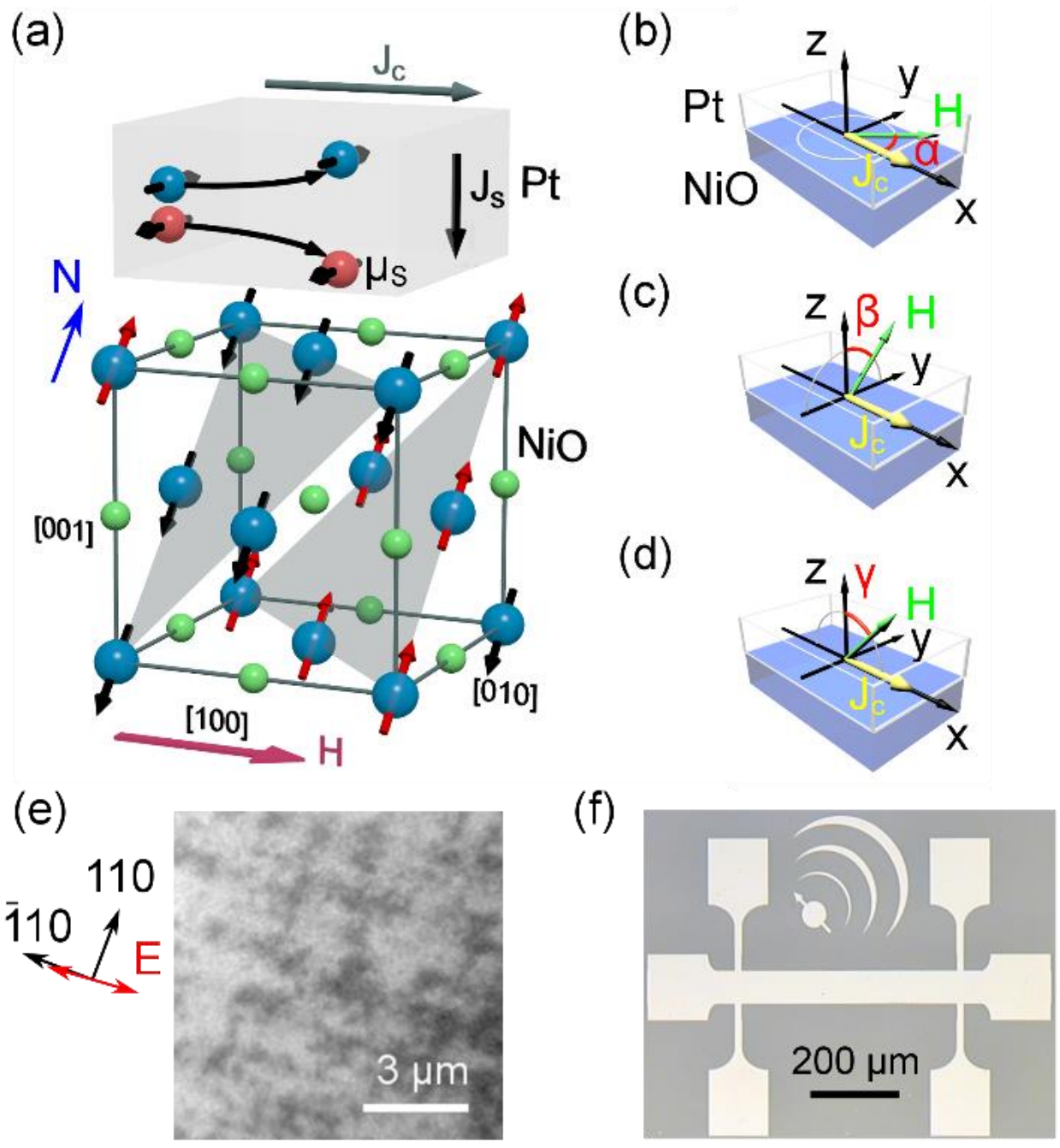

Fig. 1: (a) Measurement scheme of Spin Hall magnetoresistance (SMR) in MgO//NiO(001)/Pt. A charge current $J_c$ flows in the Pt layer, yielding a transverse spin current $J_s$ in the NiO, via the spin Hall effect. Depending on the relative orientation between the Néel vector **N** (blue arrow) and the current-induced spin polarization $\boldsymbol{\mu}_s$, the spin current polarization is modified and the spin current is converted into an additional charge current by the inverse spin Hall effect, yielding the SMR. The spin structure of one of the twelve possible $\{11\bar{2}\}$ S-domains of the NiO is shown, as well as the (111) easy plane. (b)-(d) Geometry of *xy*-, *yz*- and *xz*-planes scans, and definition of the angles α, β and γ. (e) XMLD image showing the antiferromagnetic domains in NiO at $\mu_0 H = 0$ T. The arrows indicate the orientation of the crystalline axes of the sample and of the electric field E of the X-ray beam. (f) Optical microscope image of the Hall bar design used for the magnetoresistance experiments.

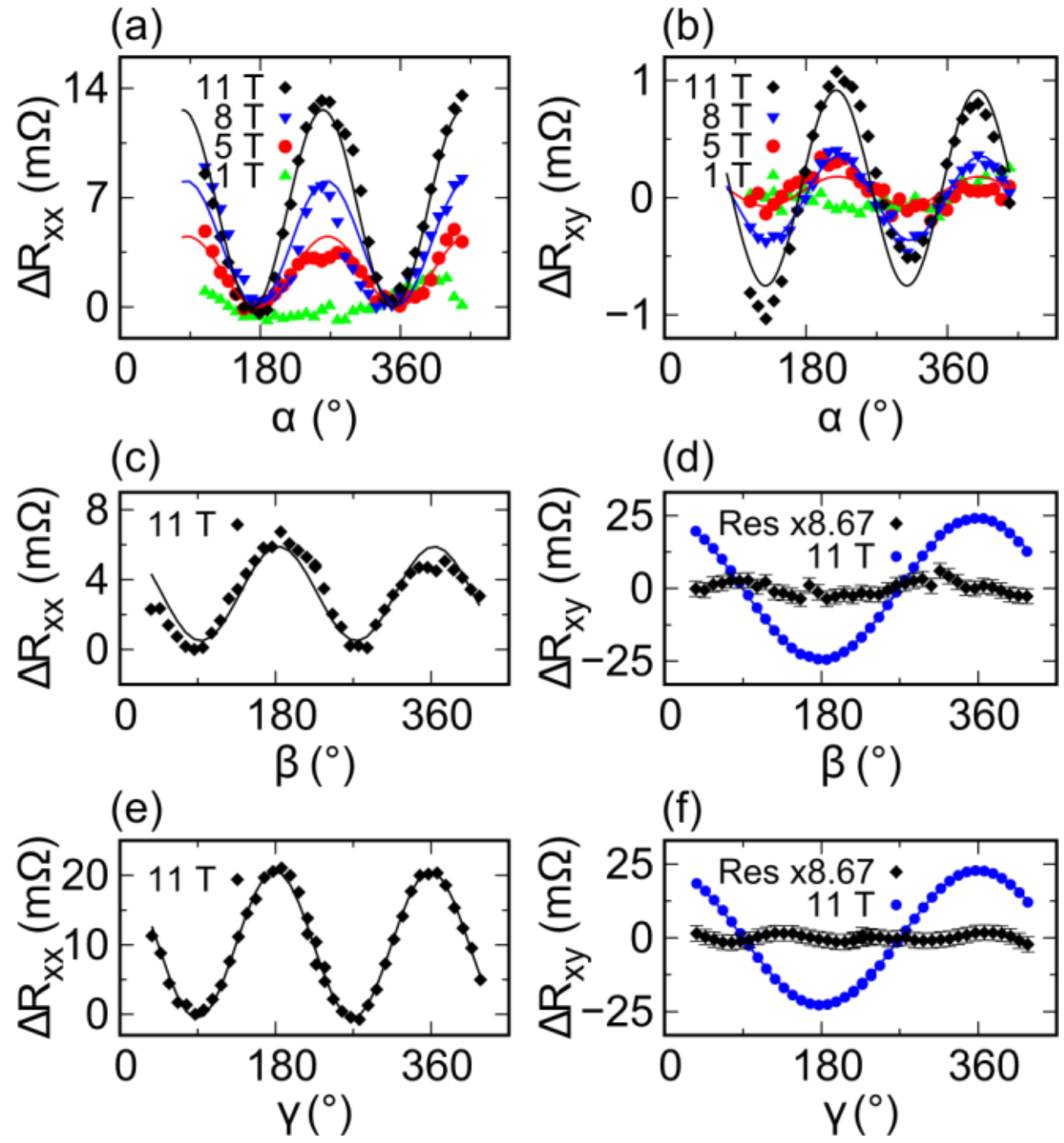

Fig. 2: Longitudinal ($\Delta R_{xx} = R_{xx}(\theta) - \min(R_{xx})$) and transverse ($\Delta R_{xy} = R_{xx}(\theta) - (\bar{R}_{xy})$) angular dependence of the resistance variation of NiO/Pt bilayers at 199.46 ± 0.02 K, where $\bar{R}$ is the angle-averaged resistance. (a)-(b) Longitudinal resistance variation $\Delta R_{xx}$ and transverse resistance variation $\Delta R_{xy}$ in the xy-plane scan, as a function of the in-plane angle α for applied field of 1, 5, 8 and 11 T. (c)-(d) $\Delta R_{xx}$ and $\Delta R_{xy}$ in the yz-plane scan, as a function of the out-of-plane angle β. (e)-(f) $\Delta R_{xx}$ and $\Delta R_{xy}$ in the xz-plane scan, as a function of the out-of-plane angle γ. The continuous lines show the $\Delta R^*\sin^2(\theta + \theta_0)$ fit of the data, except for the $\Delta R_{xy}$ data in the xz- and yz-planes scans, which is fitted by $\Delta R^*\cos(\theta + \theta_0)$. Note the predominance of the 2π-periodic Hall effect in the transverse resistance (blue circles) shown in panels d,f. The residues, shown as black squares in panels d,f, have been obtained by subtracting the fitted cosine curve to the $\Delta R_{xy}$ data and by multiplying the results by the geometrical factor length/width of the Hall bar (8.67).

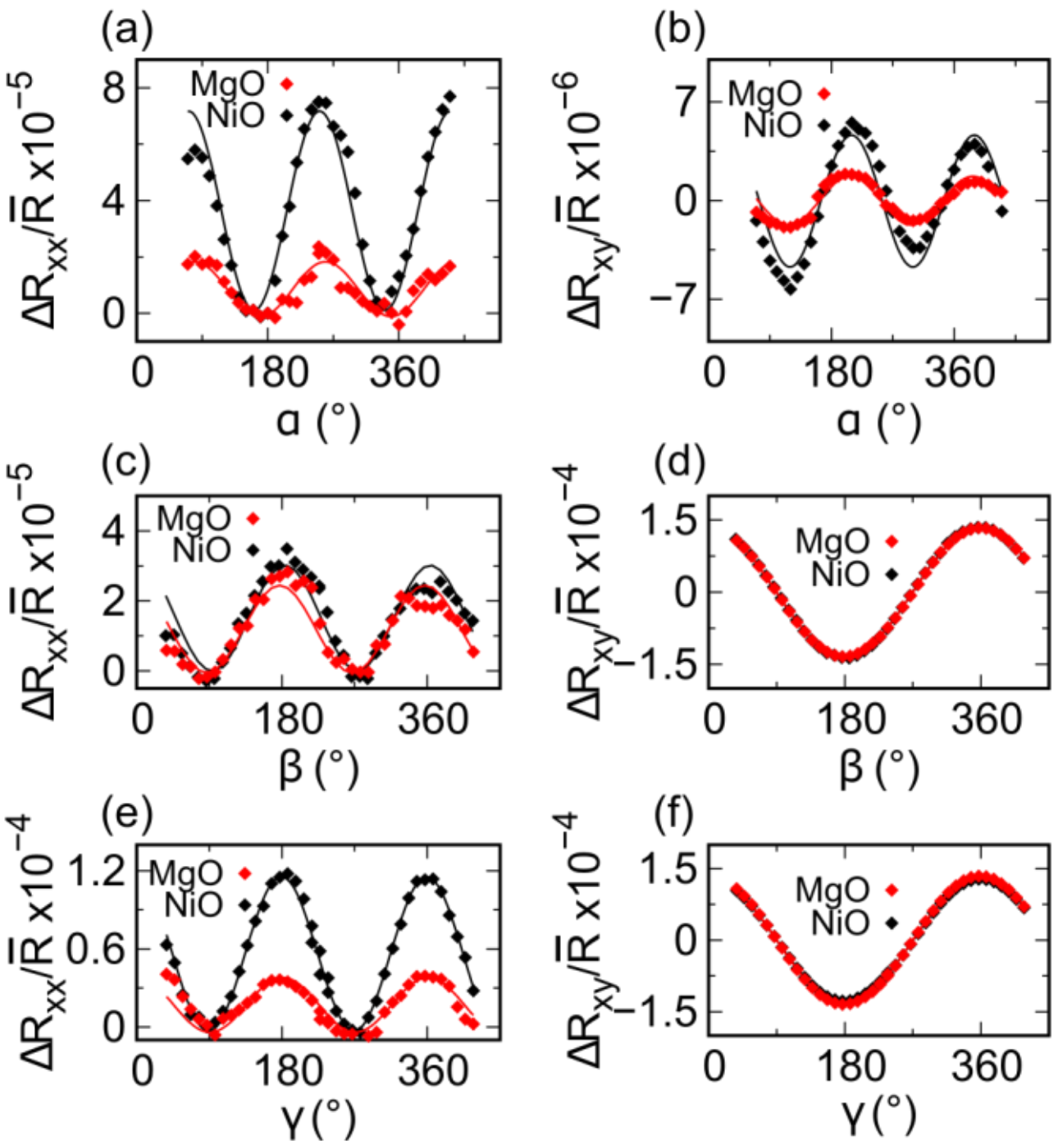

Fig. 3: Comparison between the longitudinal and transverse angular dependence of the resistance variation in MgO//NiO(90 nm)/Pt(7.5 nm), shown by black dots, and a MgO//Pt(7.5 nm) sample, shown by red dots, in an applied field at $\mu_0H$ = 11 T at 199.46 ± 0.02 K. The resistances are measured by rotating the samples in (a-b) xy-plane, (c-d) yz-plane, (e-f) xz-plane. Note that the longitudinal and transverse resistance are significantly different between the MgO and the NiO (panels a,b,e), while the longitudinal resistance in the yz-plane scan (panel c) and the Hall effect in the out of plane scans (panels d,f) are the same.

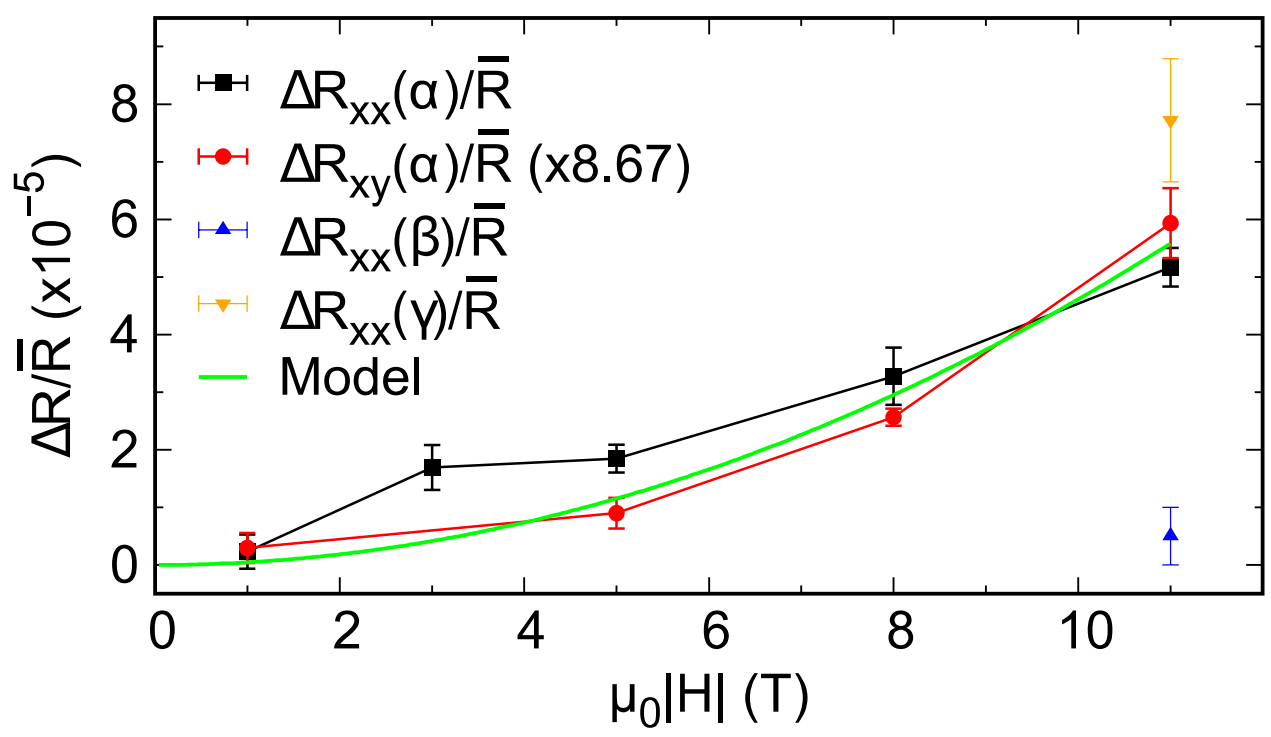

Fig. 4: Amplitude ΔR of the longitudinal and transverse resistance variation obtained by the $\Delta R*\sin^2(\theta + \theta_0)$ fit of the data for different magnetic field amplitude and divided by the angle-averaged resistance $\overline{R}$. The data were acquired in the xy-, xz- and yz-plane scans at 199.46 ± 0.02 K and are compared to the theoretical model (green solid line). No saturation is observed in the field range used up to 11 T.

| N | Domain | T-plane | $\mathbf{n}^{(0)}$ | $n_x^2$ | $n_y^2$ | $n_z^2$ | $n_x n_y$ | $n_x n_z$ | $n_y n_z$ |
|---|---|---|---|---|---|---|---|---|---|
| 1 | T1S1 | (111) | $[2\bar{1}\bar{1}]$ | 2/3 | 1/6 | 1/6 | -1/3 | -1/3 | 1/6 |
| 2 | T1S2 | | $[\bar{1}2\bar{1}]$ | 1/6 | 2/3 | 1/6 | -1/3 | 1/6 | -1/3 |
| 3 | T1S3 | | $[\bar{1}\bar{1}2]$ | 1/6 | 1/6 | 2/3 | 1/6 | -1/3 | -1/3 |
| 4 | T2S1 | $(\bar{1}11)$ | [211] | 2/3 | 1/6 | 1/6 | 1/3 | 1/3 | 1/6 |
| 5 | T2S2 | | $[12\bar{1}]$ | 1/6 | 2/3 | 1/6 | 1/3 | -1/6 | -1/3 |
| 6 | T2S3 | | $[1\bar{1}2]$ | 1/6 | 1/6 | 2/3 | -1/6 | 1/3 | -1/3 |
| 7 | T3S1 | $(1\bar{1}1)$ | $[21\bar{1}]$ | 2/3 | 1/6 | 1/6 | 1/3 | -1/3 | -1/6 |
| 8 | T3S2 | | [121] | 1/6 | 2/3 | 1/6 | 1/3 | 1/6 | 1/3 |
| 9 | T3S3 | | $[\bar{1}12]$ | 1/6 | 1/6 | 2/3 | -1/6 | -1/3 | 1/3 |
| 10 | T4S1 | $(11\bar{1})$ | $[2\bar{1}1]$ | 2/3 | 1/6 | 1/6 | -1/3 | 1/3 | -1/6 |
| 11 | T4S2 | | $[\bar{1}21]$ | 1/6 | 2/3 | 1/6 | -1/3 | -1/6 | 1/3 |
| 12 | T4S3 | | [112] | 1/6 | 1/6 | 2/3 | 1/6 | 1/3 | 1/3 |

TABLE I. Equilibrium orientation of the Néel vector in different T- and S- domains

| N | Domain | $u_{xx}$ | $u_{yy}$ | $u_{zz}$ | $u_{yz}$ | $u_{zx}$ | $u_{xy}$ |
|---|---|---|---|---|---|---|---|
| 1 | T1S1 | $-2u_1$ | $u_1$ | $u_1$ | $u_{exch} + 2u_2$ | $u_{exch} - u_2$ | $u_{exch} - u_2$ |
| 2 | T1S2 | $u_1$ | $-2u_1$ | $u_1$ | $u_{exch} - u_2$ | $u_{exch} + 2u_2$ | $u_{exch} - u_2$ |
| 3 | T1S3 | $u_1$ | $u_1$ | $-2u_1$ | $u_{exch} - u_2$ | $u_{exch} - u_2$ | $u_{exch} + 2u_2$ |
| 4 | T2S1 | $-2u_1$ | $u_1$ | $u_1$ | $u_{exch} + 2u_2$ | $-u_{exch} + u_2$ | $-u_{exch} + u_2$ |
| 5 | T2S2 | $u_1$ | $-2u_1$ | $u_1$ | $u_{exch} - u_2$ | $-u_{exch} - 2u_2$ | $-u_{exch} + u_2$ |
| 6 | T2S3 | $u_1$ | $u_1$ | $-2u_1$ | $u_{exch} - u_2$ | $-u_{exch} + u_2$ | $-u_{exch} - 2u_2$ |
| 7 | T3S1 | $-2u_1$ | $u_1$ | $u_1$ | $-u_{exch} - 2u_2$ | $u_{exch} - u_2$ | $-u_{exch} + u_2$ |
| 8 | T3S2 | $u_1$ | $-2u_1$ | $u_1$ | $-u_{exch} + u_2$ | $u_{exch} + 2u_2$ | $-u_{exch} + u_2$ |
| 9 | T3S3 | $u_1$ | $u_1$ | $-2u_1$ | $-u_{exch} + u_2$ | $u_{exch} - u_2$ | $-u_{exch} - 2u_2$ |
| 10 | T4S1 | $-2u_1$ | $u_1$ | $u_1$ | $-u_{exch} + 2u_2$ | $-u_{exch} + u_2$ | $u_{exch} - u_2$ |
| 11 | T4S2 | $u_1$ | $-2u_1$ | $u_1$ | $-u_{exch} + u_2$ | $-u_{exch} - 2u_2$ | $u_{exch} - u_2$ |
| 12 | T4S3 | $u_1$ | $u_1$ | $-2u_1$ | $-u_{exch} + u_2$ | $-u_{exch} + u_2$ | $u_{exch} + 2u_2$ |

TABLE II. Spontaneous strains (shear components) in different domains. $u_{\text{exch}} = -2.6 \cdot 10^{-3}$ is the spontaneous exchange magnetostriction which corresponds to a contraction in the [111] direction, [40] $2(u_1 + 2u_2) \equiv | u_{[2\bar{1}\bar{1}]} - u_{[0\bar{1}1]} | = 9 \cdot 10^{-5}$ is the elongation parallel to the Néel vector, the small difference $u_1 - u_2 = u_{\text{trig}} / \sqrt{2} = -1.14 \cdot 10^{-5}$ describes a distortion within the (111) plane. All numerical data [41] were determined for a bulk system.